\documentclass[onecolumn]{article}
\usepackage{amsmath}

\begin{document}

\author{Sajjad Zafar$^{\dag }$, Rizwan Ahmed$^{\ddag }$, M. Khalid Khan$%
^{\dag }$\thanks{%
mkkhan@qau.edu.pk} \\
$^{\dag }$Department of Physics, Quaid-i-Azam University \\
Islamabad 45320, Pakistan.\\
$^{\ddag }$Department of Physics and Applied Mathematics, \\
Pakistan Institute of of Engineering and Applied Sciences, \\
P. O. Nilore, Islamabad, Pakistan.}
\title{Scheme of 2-dimensional atom localization for a three-level atom via
quantum coherence}
\maketitle

\begin{abstract}
We present a scheme for two-dimensional (2D) atom localization in a
three-level atomic system. The scheme is based on quantum coherence via
classical standing wave fields between the two excited levels. Our results
show that conditional position probability is significantly phase dependent
of the applied field and frequency detuning of spontaneously emitted
photons. We obtain a single localization peak having probability close to
unity by manipulating the control parameters. The effect of atomic level
coherence on the sub-wavelength localization has also been studied. Our
scheme may be helpful in systems involving atom-field interaction.\newline
\end{abstract}

\section{Introduction}

Atom localization has been the potentially rich area of research because of
a number of important applications and fundamental research \cite%
{phil,appl1,appl2,appl3}. The idea of measuring the position of an atom has
been discussed since the beginning of quantum mechanics as proposed by
Heisenberg \cite{heisenberg} where the resolving power of equipment is
limitized by uncertainty principle. Furthermore, typical resolution of
measurement is restricted by diffraction condition where the length scale of
object to be measured must be of the order of wavelength of light used for
measurement \cite{diff}. A few of potential applications include laser
cooling and trapping of neutral atoms \cite{cool}, nano-lithography \cite%
{nano1,nano2}, atomic wavefunction measurement \cite{wave1,wave2},
Bose-Einstein condensation \cite{bose1,bose2} and coherent patterning of
matter waves \cite{matter}.

There are many existing schemes for 1D atom localization that utilizes
spatially dependent wave fields carrying advantage of atom to be localized
inside standing wave region. Earlier, the schemes were based on absorption
of light masks \cite{lightmask,lightmask1,lightmask2}, measurement of
standing wave field inside the cavity \cite%
{standing,standing1,standing2,standing3} or of atomic dipole \cite%
{dipole,dipole1,dipole2}, Raman gain process \cite{raman}, entanglement of
atom's position and internal state \cite{appl}, phase dependence of standing
wave field \cite{fazal} and the localization scheme based on resonance
fluorescence \cite{res}. Since atomic coherence and quantum interference
results in some interesting phenomenon such as Kerr non-linearity \cite%
{ker1,ker2}, four wave mixing (FWM) \cite{fwm}, electromagnetically induced
transparency (EIT) \cite{eit}, spontaneous emission enhancement or
suppression \cite{se1,se2} and optical bistability (OB) \cite{ob1,ob2}, the
idea of sub-wavelength atom localization via atomic coherence and quantum
interference has also been acquired considerable attention in recent years.

In past few years, atom localization in two dimensions has been extensively
studied for its better prospect in applications. A considerable attention
has been given to 2D atom localization in a multi-level atomic system where
two interacting orthogonal wave fields are utilized for measuring atom's
location. These studies show that 2D atom localization could be possible by
the measurement of populations of atomic states as proposed by Ivanov and
Rozhdestvensky \cite{2d} and the interaction of double-dark resonances
presented by Gao \textit{et al}. \cite{doubledark}. Other techniques
practiced for 2D atom localization includes probe absorption spectrum \cite%
{absorp,absorp1,absorp2} and controlled spontaneous emission \cite%
{spon,spon1,spon2}.

In this work, we present an efficient scheme for two-dimensional (2D) atom
localization based on quantum coherence effects in a three-level atomic
system. There exists many proposals based on the atomic coherence and
quantum interference for one-dimensional atom localization.For example, the
scheme proposed by Herkommer \textit{et al}. \cite{herk} in which they used
Autler-Townes spontaneous spectrum. In another scheme, Papalakis and Knight
\cite{knight} proposed scheme based on quantum interference that induces
sub-wavelength atom localization in three-level Lambda ($\Lambda $) system
traveling through the standing wave field. Furthermore, the scheme proposed
by Zubairy and collaborators exploited the phase of driving field for
sub-wavelength localization,hencereducing the number of localization peaks
from four to two \cite{zub2,zub21}. Gong \textit{et al}., in their schemes
demonstrated atom localization at the nodes of standing-wave field with
increased detection probability \cite{gong,gong1}. In contrast to above
schemes that discussed atom localization via single decay channel for
spontaneous emission, we have focuson fact that spontaneous emission via two
coherent decay channels may potentially improve the localization
probability. During atom-field interaction in two dimensions,the
spontaneously emitted photon carries information of the atom by creating
spatial structures of filter function at various frequencies. Typically such
spatial structures deliver lattice-like, crater-like and spike-like patterns
which are mainly dependent on dynamic interference between two orthogonal
spatially dependent fields. Subsequently,high precision in 2D atom
localization can be attained by manipulating the system parameters.

Our scheme is based on a three level-ladder configuration, where the atom is
initially prepared in the coherent superpositionof upper two excited levels.
Consequently, we studied the combined effect of relative phase between two
applied orthogonal standing wave fields and frequency of emitted photon.

The system is also discussed in the absence of atomic coherence where the
result showed radially dependence on initial atomic state preparation. We
reported multiple results for two dimensional atom localization including a
single localization peak having fairly large conditional position
probability.

This paper is organized as follows. Section II presents atomic model
followed by theoretical treatment for deriving conditional probability
distribution. In section III, we provide detailed analysis and discussion of
results regarding two-dimensional atom localization in our proposed scheme.
Finally, section IV offers concise conclusion.

\section{Model and Equations}

We consider an atom moving in $z$-direction passing through two intersecting
classical standing fields as shown in \textit{Fig. 1 (a)}. The two fields
are assumed to be orthogonal and aligned along $x$ and $y$ axis,
respectively. The internal energy levels of three level atomic system is
shown in \textit{Fig. 1(b)}.\ The two excited states $\left\vert
2\right\rangle $ and $\left\vert 1\right\rangle $\ are coupled to ground
state $\left\vert 0\right\rangle $ by vacuum modes in free space.\ Further,
the transition $\left\vert 2\right\rangle \longrightarrow \left\vert
1\right\rangle $ is driven via classical standing field with Rabi frequency $%
\Omega \left( x,y\right) $. Hence, the interaction between atom and
classical standing field is spatial dependent in $x-y$ plane.

Next we assume that the atom (with mass $m$) follows thermal distribution at
temperature $T$ so that its energy $k_{B}T$ is quite large as compared to
photon recoil $\frac{\hbar ^{2}k^{2}}{2m}.$Therefore, the atom moving along $%
z$-direction is not effected by the interaction fields and we can treat it
classically. Neglecting the kinetic energy part of Hamiltonian in Raman-Nath
approximation \cite{RN App}, the interaction Hamiltonian of the system under
dipole approximation reads as,%
\begin{equation}
H_{int}=H_{field}+H_{vacuum},  \label{Hamil}
\end{equation}%
where%
\begin{equation}
H_{field}=\Omega \left( x,y\right) e^{i\left( \Delta t+\alpha _{c}\right)
}X_{12}+\Omega ^{\ast }\left( x,y\right) e^{-i\left( \Delta t+\alpha
_{c}\right) }X_{21},
\end{equation}%
and

\begin{eqnarray}
H_{vacuum} &=&\sum\limits_{\mathbf{k}}\left( g_{\mathbf{k}}^{\left( 1\right)
}e^{-i\left( \omega _{\mathbf{k}}-\omega _{\mathbf{10}}\right) t}X_{10}b_{%
\mathbf{k}}+g_{\mathbf{k}}^{\left( 1\right) \ast }e^{i\left( \omega _{%
\mathbf{k}}-\omega _{\mathbf{10}}\right) t}X_{01}b_{\mathbf{k}}^{\dag
}\right) +  \notag \\
&&\sum\limits_{\mathbf{k}}\left( g_{\mathbf{k}}^{\left( 2\right)
}e^{-i\left( \omega _{\mathbf{k}}-\omega _{\mathbf{20}}\right) t}X_{20}b_{%
\mathbf{k}}+g_{\mathbf{k}}^{\left( 2\right) \ast }e^{i\left( \omega _{%
\mathbf{k}}-\omega _{\mathbf{20}}\right) t}X_{02}b_{\mathbf{k}}^{\dag
}\right) .
\end{eqnarray}%
Here $X_{ij}=\left\vert i\right\rangle \left\langle j\right\vert $
represents the atomic transition operator for levels $\left\vert
i\right\rangle $ and $\left\vert j\right\rangle \ $with transitions $%
\left\vert 2\right\rangle \leftrightarrow \left\vert 0\right\rangle $ and $%
\left\vert 1\right\rangle \leftrightarrow \left\vert 0\right\rangle $ are
characterized by frequencies $\omega _{\mathbf{20}}$ and $\omega _{\mathbf{10%
}}$, respectively. The frequency of coupling filed between the upper two
levels is given by $\omega _{c}$ with associated phase $\alpha _{c}$\ and
detuning parameter $\Delta =\omega _{\mathbf{c}}-\omega _{\mathbf{21}}$. The
coupling constants $g_{\mathbf{k}}^{\left( n\right) }$ $\left( n=1,2\right) $
are defined for atom-vacuum field interactions\ that corresponds to
spontaneous decay while $b_{\mathbf{k}}$ and $b_{\mathbf{k}}^{\dag }$ are
the annihilation and creation operators of vacuum modes $\mathbf{k}$.

The wave function of the whole atom-field interaction system at time $t$ can
be represented in terms of state vectors as,%
\begin{eqnarray}
\left\vert \psi \left( x,y;t\right) \right\rangle &=&\int\int dxdyf\left(
x,y\right) \left\vert x\right\rangle \left\vert y\right\rangle {\LARGE [}%
a_{2,0}\left( x,y;t\right) \left\vert 2,\left\{ 0\right\} \right\rangle
\notag \\
&&+a_{1,0}\left( x,y;t\right) \left\vert 1,\left\{ 0\right\} \right\rangle
+\sum\limits_{\mathbf{k}}a_{0,1_{\mathbf{k}}}\left( x,y;t\right) \left\vert
0,1_{\mathbf{k}}\right\rangle {\LARGE ],}  \label{state}
\end{eqnarray}%
where $f\left( x,y\right) $ is the center of mass wave function for the
atom. The position dependent probability amplitudes $a_{n,0}\left(
x,y;t\right) $ $\left( n=1,2\right) $ represents when there is no photon and
$a_{0,1_{\mathbf{k}}}\left( x,y;t\right) $ corresponds to a single photon
spontaneously emitted in $\mathbf{k}$th mode of vacuum.

In our scheme, we make use of the fact that the interaction of position
dependent Rabi frequency of atom, associated with classical standing field,
interacts with the frequency of spontaneously emitted photon \cite%
{res,herk,Zhu}. Such emitted photon carries information regrading location
of the atom. Hence, atom position measurment is subjected to the detection
of spontaneously emitted photon. Thus, the probability of atom to be located
at any position $\left( x,y\right) $ at any time $t$ can be described by
conditional position pobability distribution function $W\left( x,y\right) $
defined as,%
\begin{equation}
W\left( x,y\right) \equiv W\left( x,y;t|0,1_{\mathbf{k}}\right) =\mathcal{F}%
\left( x,y;t|0,1_{\mathbf{k}}\right) \left\vert f\left( x,y\right)
\right\vert ^{2}.  \label{prob}
\end{equation}%
Here $\mathcal{F}\left( x,y;t|0,1_{\mathbf{k}}\right) $ is the filter
function that can be defined as,%
\begin{equation}
\mathcal{F}\left( x,y;t|0,1_{\mathbf{k}}\right) =\left\vert \mathcal{N}%
\right\vert ^{2}\left\vert a_{0,1_{\mathbf{k}}}\left( x,y;t\longrightarrow
\infty \right) \right\vert ^{2},  \label{filter}
\end{equation}%
with $\mathcal{N}$ being the normalization constant. Eq.(\ref{filter}) shows
that the filter function or more generally, the conditional probabilty
distribution is provided by probability amplitude $a_{0,1_{\mathbf{k}%
}}\left( x,y;t\right) $.

In order to find the probability amplitude, we solve Schrodinger wave
equation using interaction picture Hamiltonian given by Eq.(\ref{Hamil})
with state vector as defined in Eq.(\ref{state}). The time evolution
equations of probability amplitudes are then given by,%
\begin{eqnarray}
i\dot{a}_{1,0}\left( x,y;t\right)  &=&\Omega \left( x,y\right) a_{2,0}\left(
x,y;t\right) e^{i\left( \Delta t+\alpha _{c}\right) }  \notag \\
&&+\sum\limits_{\mathbf{k}}g_{\mathbf{k}}^{\left( 1\right) }a_{0,1_{\mathbf{k%
}}}\left( x,y;t\right) e^{-i\left( \omega _{\mathbf{k}}-\omega _{\mathbf{10}%
}\right) t},  \label{eq1}
\end{eqnarray}%
\begin{eqnarray}
i\dot{a}_{2,0}\left( x,y;t\right)  &=&\Omega \left( x,y\right) a_{1,0}\left(
x,y;t\right) e^{-i\left( \Delta t+\alpha _{c}\right) }  \notag \\
&&+\sum\limits_{\mathbf{k}}g_{\mathbf{k}}^{\left( 2\right) }a_{0,1_{\mathbf{k%
}}}\left( x,y;t\right) e^{-i\left( \omega _{\mathbf{k}}-\omega _{\mathbf{20}%
}\right) t},  \label{eq2}
\end{eqnarray}%
\begin{eqnarray}
i\dot{a}_{0,1_{\mathbf{k}}}\left( x,y;t\right)  &=&g_{\mathbf{k}}^{\left(
1\right) }a_{1,0}\left( x,y;t\right) e^{i\left( \omega _{\mathbf{k}}-\omega
_{\mathbf{10}}\right) t}  \notag \\
&&+g_{\mathbf{k}}^{\left( 2\right) }a_{2,0}\left( x,y;t\right) e^{i\left(
\omega _{\mathbf{k}}-\omega _{\mathbf{20}}\right) t}.  \label{eq3}
\end{eqnarray}%
Proceeding with regular integration of Eq.(\ref{eq3}) and substituting it in
Eq.(\ref{eq1}) and (\ref{eq2}) gives us,%
\begin{eqnarray}
i\dot{a}_{1,0}\left( x,y;t\right)  &=&\Omega \left( x,y\right) a_{2,0}\left(
x,y;t\right) e^{i\left( \Delta t+\alpha _{c}\right) }  \notag \\
&&-i\int\limits_{0}^{t}dt^{\prime }a_{1,0}\left( x,y;t^{\prime }\right)
\sum\limits_{\mathbf{k}}\left\vert g_{\mathbf{k}}^{\left( 1\right)
}\right\vert ^{2}e^{-i\left( \omega _{\mathbf{k}}-\omega _{\mathbf{10}%
}\right) \left( t-t^{\prime }\right) }  \notag \\
&&-i\int\limits_{0}^{t}dt^{\prime }a_{2,0}\left( x,y;t^{\prime }\right)
\sum\limits_{\mathbf{k}}g_{\mathbf{k}}^{\left( 1\right) }g_{\mathbf{k}%
}^{\left( 2\right) }e^{\substack{ -i\omega _{\mathbf{k}}\left( t-t^{\prime
}\right) +i\omega _{\mathbf{10}}t-i\omega _{\mathbf{20}}t^{\prime }}},
\notag \\
&&
\end{eqnarray}

\begin{eqnarray}
i\dot{a}_{2,0}\left( x,y;t\right)  &=&\Omega \left( x,y\right) a_{1,0}\left(
x,y;t\right) e^{-i\left( \Delta t+\alpha _{c}\right) }  \notag \\
&&-i\int\limits_{0}^{t}dt^{\prime }a_{2,0}\left( x,y;t^{\prime }\right)
\sum\limits_{\mathbf{k}}\left\vert g_{\mathbf{k}}^{\left( 2\right)
}\right\vert ^{2}e^{-i\left( \omega _{\mathbf{k}}-\omega _{\mathbf{20}%
}\right) \left( t-t^{\prime }\right) }  \notag \\
&&-i\int\limits_{0}^{t}dt^{\prime }a_{1,0}\left( x,y;t^{\prime }\right)
\sum\limits_{\mathbf{k}}g_{\mathbf{k}}^{\left( 1\right) }g_{\mathbf{k}%
}^{\left( 2\right) }e^{\substack{ -i\omega _{\mathbf{k}}\left( t-t^{\prime
}\right) +i\omega _{\mathbf{20}}t-i\omega _{\mathbf{10}}t^{\prime }}}.
\notag \\
&&
\end{eqnarray}%
Simplifying the above two equations under Weisskopf-Wigner theory \cite{WW}
results in coupled differential equations of following form%
\begin{eqnarray}
\dot{a}_{1,0}\left( x,y;t\right)  &=&-\frac{\Gamma _{1}}{2}a_{1,0}\left(
x,y;t\right) -  \notag \\
&&\left( i\Omega \left( x,y\right) e^{i\left( \Delta t+\alpha _{c}\right) }+p%
\frac{\sqrt{\Gamma _{1}\Gamma _{2}}}{2}e^{\substack{ -i\omega _{\mathbf{21}}t
}}\right) a_{2,0}\left( x,y;t\right) ,  \label{1}
\end{eqnarray}

\begin{eqnarray}
\dot{a}_{2,0}\left( x,y;t\right) &=&-\left( i\Omega \left( x,y\right)
e^{-i\left( \Delta t+\alpha _{c}\right) }+p\frac{\sqrt{\Gamma _{1}\Gamma _{2}%
}}{2}e^{\substack{ i\omega _{\mathbf{21}}t}}\right) a_{2,0}\left(
x,y;t\right)  \notag \\
&&-\frac{\Gamma _{2}}{2}a_{2,0}\left( x,y;t\right) ,  \label{2}
\end{eqnarray}%
with $\Gamma _{1}$ and $\Gamma _{2}$ are the spontaneous decay rates for $%
\left\vert 1\right\rangle \leftrightarrow \left\vert 0\right\rangle $ and $%
\left\vert 2\right\rangle \leftrightarrow \left\vert 0\right\rangle $
transitions, respectively. These are defined as $\Gamma _{n}=2\pi \left\vert
g_{n\mathbf{k}}\right\vert ^{2}D\left( \omega _{\mathbf{k}}\right) $ where $%
D\left( \omega _{\mathbf{k}}\right) $ represents mode density at frequency $%
\omega _{\mathbf{k}}$ for vaccum. Additionally, the term $p\frac{\sqrt{%
\Gamma _{1}\Gamma _{2}}}{2}e^{\substack{ \pm i\omega _{\mathbf{21}}t}}$
corresponds to quantum interference whenever the higher energy levels in two
spontaneous emissions are very close \cite{pterm}. The parameter $p$ in our
case is defined as $p=\frac{\mathbf{\mu }_{20}.\mathbf{\mu }_{01}}{%
\left\vert \mathbf{\mu }_{20}\right\vert \left\vert \mathbf{\mu }%
_{01}\right\vert }$ which provides alignment of two matrix elements. This
clearly shows that orthogonal and parallel matrix elements are presented by $%
p=0$ and $1$, respectively. For orthogonal matrix elements i.e., $p=0$,
there is no interference and for parallel matrix elements $p=1$, the
interference is maximum. In order to solve Eq.(\ref{1}) and (\ref{2})
analytically, we assume orthogonal dipole moments that are easy to find in
nature \cite{odipole}. Further, the time dependence in above equations i.e.,
$e^{\substack{ \pm i\omega _{\mathbf{21}}t}}$ can be ignore by setting
energy difference between upper two levels $\omega _{21}$ very large as
comared to the decay rates $\Gamma _{1}$ and $\Gamma _{2}$ \cite{Gammas}.
iIntroducing the transformations as,
\begin{eqnarray}
b_{1,0}\left( x,y;t\right) &=&a_{1,0}\left( x,y;t\right) ,  \notag \\
b_{2,0}\left( x,y;t\right) &=&a_{2,0}\left( x,y;t\right) e^{i\Delta t},
\notag \\
b_{0,1_{\mathbf{k}}}\left( x,y;t\right) &=&a_{0,1_{\mathbf{k}}}\left(
x,y;t\right) ,  \label{transform}
\end{eqnarray}%
with%
\begin{equation}
\delta _{\mathbf{k}}\equiv \omega _{\mathbf{k}}-\left( \omega _{20}+\omega
_{10}\right) /2,
\end{equation}%
the rate equations (\ref{1}), (\ref{2}) and (\ref{eq3}) becomes,%
\begin{equation}
\dot{b}_{1,0}\left( x,y;t\right) =-\frac{\Gamma _{1}}{2}b_{1,0}\left(
x,y;t\right) -i\Omega \left( x,y\right) b_{2,0}\left( x,y;t\right) ,
\label{1st}
\end{equation}

\begin{eqnarray}
\dot{b}_{2,0}\left( x,y;t\right) &=&-i\Omega \left( x,y\right) e^{-i\alpha
_{c}}b_{1,0}\left( x,y;t\right)  \notag \\
&&+\left( i\Delta -\frac{\Gamma _{2}}{2}\right) b_{2,0}\left( x,y;t\right) ,
\label{2nd}
\end{eqnarray}

\begin{eqnarray}
\dot{b}_{0,1_{\mathbf{k}}}\left( x,y;t\right) &=&-ig_{\mathbf{k}}^{\left(
1\right) }b_{1,0}\left( x,y;t\right) e^{i\left( \delta _{\mathbf{k}}+\frac{%
\omega _{21}}{2}\right) t}  \notag \\
&&-ig_{\mathbf{k}}^{\left( 2\right) }b_{2,0}\left( x,y;t\right) e^{i\left(
\delta _{\mathbf{k}}-\frac{\omega _{21}}{2}-\Delta \right) t.}  \label{3rd}
\end{eqnarray}%
The set of first two differential equations can readily be solved by formal
integration techniques. At this moment, we define initial state of the
system by superposition of two orthogonal states as,%
\begin{equation}
\left\vert \psi \left( x,y;t=0\right) \right\rangle =e^{i\alpha _{p}}\sin
\left( \xi \right) \left\vert 2,\left\{ 0\right\} \right\rangle +\cos \left(
\xi \right) \left\vert 1,\left\{ 0\right\} \right\rangle ,  \label{initial}
\end{equation}%
where $\alpha _{p}$ is the phase related to pump filed. Under this initial
state with assumption that the pumping phase, $\alpha _{p}=0$ and decay
rates $\Gamma _{1}=\Gamma _{2}=\Gamma $, the solution of Eq.(\ref{1st}) and (%
\ref{2nd}) reads as,%
\begin{equation}
b_{1,0}\left( x,y;t\right) =B_{1}e^{\lambda _{1}t}+B_{1}^{\prime }e^{\lambda
_{2}t},  \label{b1}
\end{equation}%
\begin{equation}
b_{2,0}\left( x,y;t\right) =B_{2}e^{\lambda _{1}t}+B_{2}^{\prime }e^{\lambda
_{2}t},  \label{b2}
\end{equation}%
with

\begin{equation}
\lambda _{1,2}=i\frac{\Delta }{2}-\frac{\Gamma }{2}\pm \frac{i}{2}\sqrt{%
4\left( \Omega ^{2}\left( x,y\right) -\left( i\Delta -\frac{\Gamma }{2}%
\right) \frac{\Gamma }{2}\right) -\left( i\Delta -\Gamma \right) ^{2}},
\label{lambda}
\end{equation}%
\begin{equation}
B_{1}=\frac{1}{\lambda _{2}-\lambda _{1}}\left( \left( \lambda _{2}+\frac{%
\Gamma }{2}\right) \sin \left( \xi \right) +i\Omega \left( x,y\right)
e^{i\alpha _{c}}\cos \left( \xi \right) \right) ,
\end{equation}%
\begin{equation}
B_{1}^{\prime }=\frac{1}{\lambda _{1}-\lambda _{2}}\left( \left( \lambda
_{1}+\frac{\Gamma }{2}\right) \sin \left( \xi \right) +i\Omega \left(
x,y\right) e^{i\alpha _{c}}\cos \left( \xi \right) \right) ,
\end{equation}%
$\ \ \ \ \ \ \ \ \ $%
\begin{equation}
B_{2}=\frac{1}{\lambda _{2}-\lambda _{1}}\left( \left( \lambda _{2}+\frac{%
\Gamma }{2}-i\Delta \right) \sin \left( \xi \right) +i\Omega \left(
x,y\right) e^{-i\alpha _{c}}\cos \left( \xi \right) \right) ,
\end{equation}%
\begin{equation}
B_{2}^{\prime }=\frac{1}{\lambda _{1}-\lambda _{2}}\left( \left( \lambda
_{1}+\frac{\Gamma }{2}-i\Delta \right) \sin \left( \xi \right) +i\Omega
\left( x,y\right) e^{-i\alpha _{c}}\cos \left( \xi \right) \right) ,
\label{cons}
\end{equation}%
for $\lambda _{1}\neq \lambda _{2}$.

Substituting Eq.(\ref{b1}) and (\ref{b2}) in Eq.(\ref{3rd}) and following
formal integration, the probability amplitude $b_{0,1_{\mathbf{k}}}\left(
x,y;t\right) $ for interaction time much larger than the decay rates $\left(
\Gamma _{1}t,\Gamma _{2}t\gg 1\right) $ offers,%
\begin{eqnarray}
b_{0,1_{\mathbf{k}}}\left( x,y;t\rightarrow \infty \right)  &=&g_{\mathbf{k}%
}^{\left( 1\right) }\left( \tfrac{B_{1}}{\delta _{\mathbf{k}}+\frac{\omega
_{21}}{2}-i\lambda _{1}}+\tfrac{B_{1}^{\prime }}{\delta _{\mathbf{k}}+\frac{%
\omega _{21}}{2}-i\lambda _{2}}\right) +  \notag \\
&&g_{\mathbf{k}}^{\left( 2\right) }\left( \tfrac{B_{2}}{\delta _{\mathbf{k}}-%
\frac{\omega _{21}}{2}-\Delta -i\lambda _{1}}+\tfrac{B_{2}^{\prime }}{\delta
_{\mathbf{k}}-\frac{\omega _{21}}{2}-\Delta -i\lambda _{2}}\right) .
\end{eqnarray}%
Incorporating the constants as specified in Eq.(\ref{lambda})-(\ref{cons})
with trivial simplifications, we get%
\begin{eqnarray}
b_{0,1_{\mathbf{k}}}\left( x,y;t\rightarrow \infty \right)  &=&\frac{g_{%
\mathbf{k}}^{\left( 1\right) }}{2}\left( \tfrac{\sin \left( \xi \right)
-e^{i\alpha _{c}}\cos \left( \xi \right) }{\delta _{\mathbf{k}}+\frac{\omega
_{21}}{2}+\Omega \left( x,y\right) +i\frac{\Gamma }{2}}+\tfrac{\sin \left(
\xi \right) +e^{i\alpha _{c}}\cos \left( \xi \right) }{\delta _{\mathbf{k}}+%
\frac{\omega _{21}}{2}-\Omega \left( x,y\right) +i\frac{\Gamma }{2}}\right) +
\notag \\
&&\frac{g_{\mathbf{k}}^{\left( 2\right) }}{2}\left( \tfrac{\cos \left( \xi
\right) -e^{-i\alpha _{c}}\sin \left( \xi \right) }{\delta _{\mathbf{k}}-%
\frac{\omega _{21}}{2}-\Delta +\Omega \left( x,y\right) +i\frac{\Gamma }{2}}+%
\tfrac{\cos \left( \xi \right) +e^{-i\alpha _{c}}\sin \left( \xi \right) }{%
\delta _{\mathbf{k}}-\frac{\omega _{21}}{2}-\Delta -\Omega \left( x,y\right)
+i\frac{\Gamma }{2}}\right) .  \notag \\
&&
\end{eqnarray}%
Consequently, the required conditional probability distribution of finding
the atom in state $\left\vert 0\right\rangle $ with emitted photon of
frequency $\omega _{\mathbf{k}}$ corresponding to reservoir mode $\mathbf{k}$
is depicted by%
\begin{equation}
W\left( x,y\right) =\left\vert \mathcal{N}\right\vert ^{2}\left\vert f\left(
x,y\right) \right\vert ^{2}\left\vert b_{0,1_{\mathbf{k}}}\left(
x,y;t\rightarrow \infty \right) \right\vert ^{2},  \label{w}
\end{equation}%
where we have used the transformation $b_{0,1_{\mathbf{k}}}\left(
x,y;t\right) =a_{0,1_{\mathbf{k}}}\left( x,y;t\right) $ from Eq. (\ref%
{transform}). Since the center-of-mass wave function $f(x,y)$ for atom is
assumed to be almost constant over many wavelengths of the standing-wave
fields in $x-y$ plane, the conditional probability distribution $W\left(
x,y\right) $ for atom localization is determined by filter function as
defined in Eq. (\ref{filter}).

\section{Numerical Results and Discussion}

In this section, we will discuss the conditional probability distribution of
an atom employing few numerical results based on filter function $\mathcal{F}%
(x,y)$. We will then swing the system parameter to show how atom
localization can be attained via quantum coherence. In our analysis, we have
considered two orthogonal standing waves with corresponding Rabi frequency $%
\Omega \left( x,y\right) $ $=\Omega _{1}\sin \left( k_{1}x\right) +\Omega
_{2}\sin \left( k_{2}y\right) $ \cite{absorp1}. Further, all the parameters
are taken in terms of decay rate $\Gamma $. Apparently, the filter function $%
\mathcal{F}(x,y)$ depends on the paramaters of standing wave driving fields
and frequency of the emitted photon, rather it also depends on the
interference effects \cite{absorp1}. Since the two spontaneous decay
channels $\left\vert 1\right\rangle \rightarrow \left\vert 0\right\rangle $
and $\left\vert 2\right\rangle \rightarrow \left\vert 0\right\rangle $
interact via same vaccum modes, quantum interfence subsists. However, we
have neglected quantum interference by setting paramater $p=0$ in our
analysis. Moreover, the dynamically induced interference due to two
orthogonal standing wave fields does play a considerable effect. Hence, atom
localization in 2D can be manipulated by various parameters. As the filter
function $\mathcal{F}(x,y)$ depends on $\Omega \left( x,y\right) $ which
itself comprises of $\sin \left( k_{1}x\right) $ and $\sin \left(
k_{2}y\right) $, the localization is possible for only those values of $%
(x,y) $ for which $\mathcal{F}(x,y)$ reveals maxima. Here, the analytical
form of $\mathcal{F}(x,y)$ appears to be quite cumbersome, so we will
provide only numerical results for precise atom location in two dimensions.

Since the atom is initially prepared in the superposition of upper two
excited states $\left\vert 1\right\rangle $ and $\left\vert 2\right\rangle $
via Eq.(\ref{initial}), the state is strongly dependent on coupling phase
for upper two levels $\left\vert 1\right\rangle $ and $\left\vert
2\right\rangle $ i.e., $\alpha _{c}$ which shows quantum coherence is phase
dependent. Accordingly, we have considered three values of phase that is $%
\alpha _{c}=0,\frac{\pi }{2}$ and $\pi $.

In case of $\alpha _{c}=$ $\frac{\pi }{2}$, we first provide conditional
probability distribution by plotting filter function $\mathcal{F}(x,y)$ as a
function of $(k_{1}x,k_{2}y)$ over a single wavelength for different value
of detuning of spontaneously emitted photon i.e., $\delta _{\mathbf{k}}$
\cite{result1-0}. From \textit{Fig. 2(a)-(d)}, it is evident that the filter
function $\mathcal{F}(x,y)$ strongly depends on the detuning of
spontaneously emitted photon. When $\delta _{\mathbf{k}}=9.3\Gamma $, the
location of atom is distributed in all four quadrants in $x-y$ plane, as
shown in \textit{Fig. 2(a)}. Switching values to $\delta _{\mathbf{k}%
}=5.3\Gamma $ , the location is restricted to quadrant I and IV with peaks
providing crater like pattern [\textit{Fig. 2(b)}]. On further refining the
detuning to $\delta _{\mathbf{k}}=2.9\Gamma $, the peaks get narrowed, as
depicted in \textit{Fig. 2(c)}. Furthermore, \textit{Fig. 2(d)} illustrate
the effect of detuning set to appropriate value i.e., $\delta _{\mathbf{k}%
}=0.1\Gamma $. The filter function $\mathcal{F}(x,y)$ furnishes two spike
like patterns corresponding to maximas located in quadrant I and IV at $%
(k_{1}x,k_{2}y)=(\frac{\pi }{2},\frac{\pi }{2})$ and $(-\frac{\pi }{2},-%
\frac{\pi }{2})$ in $x-y$ plane which clearly indicates that high precision
and localization in two dimension can be obtained when emitted photon is
nearly resonant with the corresponding atomic transition. Consequently, the
probability of finding the atom at each location is $\frac{1}{2}$ which is
twice as compared to the probability obtained in earlier cases \cite%
{2d,absorp,spon2,result1-2}.

In \textit{Fig. 3}, we plot the filter function $\mathcal{F}(x,y)$ versus $%
(k_{1}x,k_{2}y)$ by modulating the detuning of spontaneously emitted photon
in 2D for $\alpha _{c}=$ $0$. \textit{Fig. 3(a)} illustrates the results
when detuning is large i.e., $\delta _{\mathbf{k}}=12.4\Gamma $. The lattice
like structure obtained gives distributed on the diagonal in II and IV
quadrants. This specifies that the atom localization peaks are determined by
$k_{1}x+k_{2}y=2p\pi $ or $k_{1}x+k_{2}y=2\left( q+1\right) \pi $ where $p$
and $q$ are intergers. Refining $\delta _{\mathbf{k}}$ to $9.5\Gamma ,$ the
position probability of atom is rather complicated due to the interference
of the two fields and the filter function $\mathcal{F}(x,y)$ is dispersed in
quadrant II, III and IV. However, the distribation is mainly localized in
quardant III, as presented in \textit{Fig. 3(b)}. Narrowing the detuning
parameter to $\delta _{\mathbf{k}}=6.0\Gamma $, the location is distributed
in quadrant III with a crater like structure, as shown in \textit{Fig. 3(b)}%
. Such crater like structure persists in quardant III for $\delta _{\mathbf{k%
}}\in \left[ 5.2\Gamma ,6.9\Gamma \right] $, offering atom localized at the
circle [\textit{Fig. 3(c)}]. Tuning the photon detuning to $\delta _{\mathbf{%
k}}=2.4\Gamma $, a single spike is achieved at $(-\frac{\pi }{2},-\frac{\pi
}{2})$ as illustrated in \textit{Fig. 3(d)} which indicates that probability
of finding the atom within single wavelegth in 2D, is increased by a factor
of 2 (as in \textit{Fig.2(d)}). Hence, we can say that atom localization is
undeniably acquired in 2D.

From Eq. (\ref{w}) with $\Omega \left( x,y\right) $ $=\Omega _{1}\sin \left(
k_{1}x\right) +\Omega _{2}\sin \left( k_{2}y\right) $, we can easily
identify that the filter function $\mathcal{F}(x,y)$ remains unaltered under
transform $0\leftrightarrow \pi $ and $(k_{1}x,k_{2}y)\leftrightarrow
(-k_{1}x,-k_{2}y).$ Therefore, $\mathcal{F}(x,y;\alpha _{c}=0)=\mathcal{F}%
(-x,-y;\alpha _{c}=\pi )$ and we obtain vice versa results as in previous
case $(\alpha _{c}=0)$ \ However, the localization distribution and peak are
shifted in quadrant I, as shown in \textit{Fig. 4(a)-(d)}.

These results identify the strong association of detuning of spontaneously
emitted photon to the localization of atom. Furthermore, the peaks of atom
localization in all of the above cases are obtained at antinodes of the
standing fields with precise localization destroyed for large frequency of
the spontaneously emitted photon. Indeed, the localization seems to be
possible when emitted photon are near in resonance to the atomic transitions.

Finally we present the significance of initial conditions on atom
localization by explicitly preparing the atomic system is single state.
Therefore for $\alpha _{c}=0$, and setting $\xi =0$ in Eq. (\ref{initial}),
the system is initially in state $\left\vert 1\right\rangle $ which provides
the distribution of localization peaks takes place in quadrant I and III as
depicted in \textit{Fig. 5(a)}. Hence, the probability of finding the atom
at a single location, in 2D is decreased. Therefore, the number of peaks
increased by a factor of 2,\ as compared to the case when atom is initially
prepared in superposition of states $\left\vert 1\right\rangle $ and $%
\left\vert 2\right\rangle $ [see \textit{Fig. 3(d)}]. The reason behind is
the absence of atomic coherence for two decaying states $\left\vert
1\right\rangle $ and $\left\vert 2\right\rangle $. A similar result can also
be obtained for $\xi =\pi $ where again the probability is decreased by a
factor of 2 as shown in \textit{Fig. 4(d)}. However, \textit{Fig. 5(b})
shows sharp peaks by setting $\xi =\frac{\pi }{2}$. This indeed provides us
high resolution and precision for atomic localization in 2D in the absence
of atomic coherence.

\section{Conclusions}

In summary, we have proposed and analyzed atom localization for a three
level atomic system in two dimensions. The scheme under consideration is
based on the phenomenon of spontaneous emission when the atom interacts with
spatially dependent standing orthogonal fields. Following the position
dependent atom-field interaction, the precise location of atom in 2D, can be
achieved by detecting frequency of spontaneously emitted photon.
Consequently, the interaction provides various structures of filter function
such as lattice like structure, crater like structure and most importantly,
the localization spikes. The phenomenon of quantum coherence originates from
coupling of two excited levels to standing wave fields. Our results shows
that not only the relative phase between two orthogonal standing wave fields
but also the frequency detuning substantially controls florescence spectra
in conditional probaility distribution. The localization pattern generates a
single spike for inphase position dependent fields. However, the
localization pattern is destroyed with an increase in frequency detuning.
Remarkably, the pattern of localization peaks remains unaltered with varying
vacuum field detuning which is the major advantage of our scheme. We have
also presented the effect of initial state prepration on atom localization.
In the absence of atomic coherence, the localization probability decreases
with increase in spatial resolution. Our analysis indeed provides efficient
way for atom localization in two dimension that may be productive for laser
cooling and atom nano-lithography \cite{result1-2}.\newline

\section{Figure Captions}

\textbf{Fig-1:} Schematic diagram of the system. $\left( a\right) $ Atom
moving along $z-$axis interacts with two dimensional position dependent
field in $x-y$ plane; $\left( b\right) $ Atomic model under consideration.
The two excited levels $\left\vert 1\right\rangle $\ and $\left\vert
2\right\rangle $\ are coupled by two dimensional standing wavefield $\Omega
\left( x,y\right) $\ with level $\left\vert 2\right\rangle $\ having a
finite detuning $\Delta $. Both the excited levels $\left\vert
1\right\rangle $\ and $\left\vert 2\right\rangle $ decay spontaneously to
ground state via decay rates $\Gamma _{1}$ and $\Gamma _{2}$, respectively

\textbf{Fig-2:} Filter function $\mathcal{F}(x,y)$ as a function of $%
(k_{1}x,k_{2}y)$ for detuning of spontaneously emitted photon, $\left(
a\right) $ $\delta _{\mathbf{k}}=9.3$; $\left( b\right) $ $\delta _{\mathbf{k%
}}=5.3$; $\left( c\right) $ $\delta _{\mathbf{k}}=2.9$; $\left( d\right) $ $%
\delta _{\mathbf{k}}=0.1$. Other parameters are $\alpha _{c}=\frac{\pi }{2}$%
, $\xi =\frac{\pi }{4}$, $\Delta =2.5$, $\omega _{21}=20$ and $\Omega
_{1}=\Omega _{2}=5.$All system parameters are scaled in units of $\Gamma $.

\textbf{Fig-3:} Filter function $\mathcal{F}(x,y)$ as a function of $%
(k_{1}x,k_{2}y)$ for detuning of spontaneously emitted photon, $\left(
a\right) $ $\delta _{\mathbf{k}}=12.4$; $\left( b\right) $ $\delta _{\mathbf{%
k}}=9.5$; $\left( c\right) $ $\delta _{\mathbf{k}}=6.0$; $\left( d\right) $ $%
\delta _{\mathbf{k}}=2.4$. Other parameters are $\alpha _{c}=0$, $\xi =\frac{%
\pi }{4}$, $\Delta =2.5$, $\omega _{21}=20$ and $\Omega _{1}=\Omega _{2}=5.$%
All system parameters are scaled in units of $\Gamma $.

\textbf{Fig-4:} Filter function $\mathcal{F}(x,y)$ as a function of $%
(k_{1}x,k_{2}y)$ for detuning of spontaneously emitted photon, $\left(
a\right) $ $\delta _{\mathbf{k}}=12.4$; $\left( b\right) $ $\delta _{\mathbf{%
k}}=9.5$; $\left( c\right) $ $\delta _{\mathbf{k}}=6.0$; $\left( d\right) $ $%
\delta _{\mathbf{k}}=2.4$. Other parameters are $\alpha _{c}=\pi $, $\xi =%
\frac{\pi }{4}$, $\Delta =2.5$, $\omega _{21}=20$ and $\Omega _{1}=\Omega
_{2}=5.$All system parameters are scaled in units of $\Gamma $.

\textbf{Fig-5:} Filter function $\mathcal{F}(x,y)$ as a function of $%
(k_{1}x,k_{2}y)$ for detuning of spontaneously emitted photon, $\left(
a\right) $ $\xi =0$; $\left( b\right) $ $\xi =\frac{\pi }{2}$. Other
parameters are $\alpha _{c}=\pi $, $\delta _{\mathbf{k}}=2.4$, $\Delta =2.5$%
, $\omega _{21}=20$ and $\Omega _{1}=\Omega _{2}=5.$All system parameters
are scaled in units of $\Gamma $.

\end{document}